\documentclass[preprintnumbers, prd, twocolumn, showpacs, floatfix,
preprintnumbers, letterpaper, superscriptaddress,nofootinbib]{revtex4}
\usepackage{graphicx,epsfig}
\usepackage{amsmath}
\usepackage {amssymb}
\usepackage{subfigure}

\usepackage{relsize}

\newcommand{\be}{\begin{equation}}
\newcommand{\ee}{\end{equation}}
\newcommand{\bea}{\begin{eqnarray}}
\newcommand{\eea}{\end{eqnarray}}
\newcommand{\nn}{\nonumber}

\begin{document}

\title{Cosmological applications of $F(T,T_G)$ gravity}

\author{Georgios Kofinas}
\email{gkofinas@aegean.gr} \affiliation{Research Group of Geometry,
Dynamical Systems and Cosmology,
Department of Information and Communication Systems Engineering\\
University of the Aegean, Karlovassi 83200, Samos, Greece}

\author{Emmanuel N. Saridakis}
\email{Emmanuel_Saridakis@baylor.edu}
\affiliation{Physics Division,
National Technical University of Athens, 15780 Zografou Campus,
Athens, Greece} \affiliation{Instituto de F\'{\i}sica, Pontificia
Universidad de Cat\'olica de Valpara\'{\i}so, Casilla 4950,
Valpara\'{\i}so, Chile}


\begin{abstract}
We investigate the cosmological applications of $F(T,T_G)$ gravity,
which is a novel modified gravitational theory based on the torsion
invariant $T$ and the teleparallel equivalent of the Gauss-Bonnet
term $T_{G}$. $F(T,T_{G})$ gravity differs from both $F(T)$ theories
as well as from $F(R,G)$ class of curvature modified gravity, and
thus its corresponding cosmology proves to be very interesting. In
particular, it provides a unified description of the cosmological
history  from early-times inflation to late-times self-acceleration,
without the inclusion of a cosmological constant. Moreover, the dark
energy  equation-of-state parameter can be quintessence or
phantom-like, or experience the phantom-divide crossing, depending
on the parameters of the model.
\end{abstract}


\maketitle

\section{Introduction}
\label{Introduction}

Since theoretical arguments and observational data suggest that the universe
passed through an early-times inflationary stage and resulted in a late-times
accelerated phase, a large amount of research was devoted to explain this behavior.
In general, one can follow two ways to achieve it. The first direction is to
alter the universe content by introducing additional fields,
canonical scalar, phantom scalar, both scalars, vector fields etc, that is
introducing the concepts of the inflaton and/or the dark energy, which can be
extended in a huge class of models (see
\cite{Bassett:2005xm,Copeland:2006wr,Cai:2009zp} and references
therein). The second way is to modify the
gravitational sector instead (see \cite{Capozziello:2011et} and
references therein). Note however that one can in principle transform from
one approach to the other, since the important point is the number of
degrees of freedom beyond standard model particles and General Relativity (GR)
\cite{Sahni:2006pa}.

In modified gravitational theories one usually extends the curvature-based
Einstein-Hilbert action. However, a different and interesting class
of gravitational modification arises when one extends the action of the
equivalent torsional formulation of General Relativity. In particular, since
Einstein's years it was known that one can construct the so-called
``Teleparallel Equivalent of General Relativity'' (TEGR)
\cite{ein28,Hayashi:1979qx,Maluf,Pereira,Maluf:2013gaa}, that is attributing
gravity to torsion instead of curvature, by using instead of the torsion-less
Levi-Civita connection the curvature-less Weitzenb{\"{o}}ck one. In such a
formulation the gravitational Lagrangian results from contractions of the
torsion tensor and is called the ``torsion scalar'' $T$, similarly to the
General Relativity Lagrangian, i.e. the ``curvature scalar'' $R$, which is
constructed by contractions of the curvature  tensor. Hence, similarly to
the $f(R)$ extensions of General Relativity
\cite{DeFelice:2010aj,Nojiri:2010wj}, one can construct $f(T)$ extensions of
TEGR  \cite{Ferraro:2006jd,Linder:2010py}. The interesting feature in
this extension is that  $f(T)$ does not coincide with $f(R)$ gravity,
despite the fact that TEGR coincides with General Relativity. Since it is a
new gravitational modification class, its corresponding cosmological behavior
and black hole solutions have been studied in detail
\cite{Ferraro:2006jd,Linder:2010py,Chen:2010va,Myrzakulov:2010vz,Li:2011wu,Wu:2010mn,Wang:2011xf}.

However, apart from the simple modifications of curvature gravity, one can
construct more complicated actions introducing higher-curvature corrections
such as the Gauss-Bonnet combination $G$
\cite{Wheeler:1985nh,Antoniadis:1993jc} or arbitrary
functions $f(G)$
  \cite{Nojiri:2005jg,DeFelice:2008wz,Davis:2007ta}, Weyl combinations
\cite{Mannheim:1988dj}, Lovelock
combinations
\cite{Lovelock:1971yv,Deruelle:1989fj} etc. Hence, one can
follow the same direction starting from the teleparallel formulation of
gravity, and construct actions involving higher-torsion corrections. Indeed,
in our recent work \cite{Kofinas:2014owa} we first constructed the
teleparallel equivalent of the Gauss-Bonnet term $T_G$ (which is a new
quartic torsional scalar which reduces to a topological invariant in four dimensions), and
then, using also the torsion scalar $T$, we constructed $F(T,T_G)$ gravity
(see
\cite{muller-hoissen,zanelli,Harko:2014sja} for different constructions of
torsional actions). This is a new class of gravitational
modification, since it is different from both $f(T)$ gravity as well as from
$f(R,G)$ gravity.

Since $F(T,T_G)$ gravity is a novel modified gravity theory, in the present
work we are interested in investigating its cosmological applications. In
particular, after extracting the Friedmann equations, we define the
effective dark energy sector and the various observables such as the density
parameters and the dark energy equation-of-state parameter. Then,
considering specific $F(T,T_G)$  ansatzes we investigate the inflation
realization and the late-times acceleration. The plan of the work is as
follows: In section \ref{model} we review $F(T,T_G)$ gravity. In section
\ref{Fcosmo} we apply it in a cosmological framework, extracting the
corresponding equations and defining the various observables, while in
section \ref{speccases} we analyze some specific cases. Finally, section
\ref{Conclusions} is devoted to the conclusions.

\section{$F(T,T_G)$ gravity}
\label{model}

Let us give a brief review of $F(T,T_G)$ gravity \cite{Kofinas:2014owa}. Although in this manuscript
we are interested in its cosmological application, we will present the formulation in $D$-dimensions
where it is non-trivial, and then discuss $F(T,T_G)$ gravity in $D=4$. In the teleparallel
formulation of gravity theories, the dynamical variables are the
vielbein field $e_a(x^\mu)$ and the connection 1-forms $\omega^a_{\,\,\,
b}(x^\mu)$ which defines the parallel transportation\footnote{In this
manuscript the notation is as follows: Greek indices $\mu, \nu,$...
run over all space-time coordinates, while Latin indices $a, b, $... run
over the tangent
space.}. We can express them in components in terms of coordinates as
$e_a=e^{\,\,\, \mu}_a\partial_\mu$ and $\omega^a_{\,\,\,b}=\omega^a_{\,\,\,
b\mu}dx^\mu=\omega^a_{\,\,\,bc}e^c$, while we define the dual vielbein as
$e^a=e^a_{\,\,\, \mu}d x^\mu$.
The vielbein commutation relations read
\begin{equation}
[e_{a},e_{b}]=C^{c}_{\,\,\,ab}e_{c}\,,
\label{ghw}
\end{equation}
where  the structure coefficients functions $C^{c}_{\,\,\,ab}$ are written as
\begin{equation}
C^{c}_{\,\,\,ab}=e_{a}^{\,\,\,\mu}
e_{b}^{\,\,\,\nu}(e^{c}_{\,\,\,\mu,\nu}-e^{c}_{\,\,\,\nu,\mu})
\label{structurefun}\,,
\end{equation}
with a comma denoting differentiation. Thus, we can define the torsion
tensor, expressed in tangent components  as
\begin{equation}
T^{a}_{\,\,\,bc}=
\omega^{a}_{\,\,\,cb}-\omega^{a}_{\,\,\,bc}-C^{a}_{\,\,\,bc}\,,
\end{equation}
 while the curvature tensor is
\begin{equation}
R^{a}_{\,\,\,bcd}\!=\omega^{a}_{\,\,\,bd,c}-\omega^{a}_{\,\,\,
bc,d}+\omega^{e}_{\,\,\,bd}\omega^{a}_{\,\,\,ec}
-\omega^{e}_{\,\,\,bc}\omega^{a}_{\,\,\,ed}  -C^{e}_{\,\,\, cd}
\omega^{a}_{\,\,\, be}.
\label{curvaturebastard}
\end{equation}
 Furthermore, we use  the metric tensor $g$ to make the vielbein orthonormal
$g(e_a,e_b)=\eta_{ab}$, where
$\eta_{ab}=\text{diag}(-1,1,...1)$, and thus we obtain the useful
relation
\begin{equation}
\label{metrdef}
g_{\mu\nu} =\eta_{ab}\, e^a_{\,\,\,\mu}  \, e^b_{\,\,\,\nu},
\end{equation}
and indices $a,b,...$ are raised/lowered with the Minkowski metric
$\eta_{ab}$.
 Finally, it proves convenient to define the  contorsion tensor as
\begin{equation}
\mathcal{K}_{abc}=\frac{1}{2}(T_{cab}-T_{bca}-T_{abc}
)=-\mathcal{K}_{bac}.
\end{equation}

We now impose the condition of teleparallelism, namely $R_{\,\,\,bcd}^{a}=0$,
 which holds in all frames. One way to realize this condition is by assuming
the Weitzenb{\"{o}}ck connection $\tilde{\omega}_{\,\,\,\mu\nu}^{\lambda}$
 defined in terms of the vielbein in
all coordinate frames as
$\tilde{\omega}_{\,\,\,\mu\nu}^{\lambda}=e_{a}^{\,\,\,\lambda}e^{a}_{\,\,\,
\mu
, \nu }$, or expressed in the preferred tangent-space components
as $\tilde{\omega}_{\,\,\,bc}^{a}=0$. The Ricci scalar
$\bar{R}$ corresponding to the usual Levi-Civita connection can be expressed
as \cite{Pereira,Maluf}
\begin{equation}
e\bar{R}=-eT+2(eT_{\nu}^{\,\,\,\nu\mu})_{,\mu}\,,
\label{ricciscalar}
\end{equation}
where we have defined the ``torsion scalar'' $T$ as
\begin{eqnarray}
T&=&\frac{1}{4}T^{\mu\nu\lambda}T_{\mu\nu\lambda}+\frac{1}{2}T^{\mu\nu\lambda
}
T_{\lambda\nu\mu}-T_{\nu}^{\,\,\,\nu\mu}T^{\lambda}_{\,\,\,\lambda\mu},
\label{Tscalar}
\end{eqnarray}
and  $e=\det{(e^{a}_{\,\,\,\mu})}=\sqrt{|g|}$.

One can now clearly see that the Lagrangian density $e\bar{R}$ of General
Relativity, that is the one calculated with the Levi-Civita connection,
and the torsional density $-eT$ differ only by a total derivative.
Hence, the Einstein-Hilbert action
\begin{eqnarray}
S_{EH}=\frac{1}{2\kappa_{D}^2}\int_{M}d^{D}\!x\,e\,\bar{R},
\label{GenRelaction}
\end{eqnarray}
up to boundary terms is equivalent to the action
\begin{eqnarray}
S_{tel}^{(1)}
&\!\!=\!\!&-\frac{1}{2\kappa_{D}^{2}}\int_{M}\!\!d^{D}\!x\,
e\,T
\label{teleaction}
\end{eqnarray}
in the sense that varying (\ref{GenRelaction}) with respect to the metric
and varying (\ref{teleaction}) with respect to the vielbein gives rise to the
same equations of motion ($\kappa_{D}^2$ is the $D$-dimensional gravitational constant)
\cite{Maluf:2013gaa}. That is why the above theory, where one
uses torsion to describe the gravitational field and imposes the
teleparallelism condition, was dubbed by Einstein as Teleparallel Equivalent
of General Relativity (TEGR).

The recipe of the construction of TEGR was to express the Ricci
scalar $R$ corresponding to a general connection as the Ricci scalar
$\bar{R}$ calculated with the Levi-Civita connection, plus terms arising from
the torsion tensor. Then, by imposing the teleparallelism condition
$R^{a}_{\,\,\,bcd}=0$, we obtained that $\bar{R}$ is equal to a
torsion scalar plus a total derivative. Hence, we can follow the same
steps, but using the  Gauss-Bonnet combination
$G=R^{2}-4R_{\mu\nu}R^{\mu\nu}+R_{\mu\nu\kappa\lambda}R^{\mu\nu\kappa\lambda}
$  instead of the Ricci scalar. In \cite{Kofinas:2014owa} we have derived the
teleparallel equivalent of Gauss-Bonnet gravity characterized by the new torsion scalar $T_{G}$,
as well as the equations of motion of the modified gravity defined by the function $F(T,T_{G})$.
In the following, we give
the corresponding expressions when restricted to the Weitzenb{\"{o}}ck connection
$\omega^{a}_{\,\,\,bc}=0$
(the tildes are withdrawn for notational simplicity). It is
\begin{equation}
e\bar{G}
\!=\!eT_{G}\!+\!\text{total diverg.},
\label{TG}
\end{equation}
where $\bar{G}$ is the Gauss-Bonnet term calculated by the Levi-Civita
connection, and
\begin{eqnarray}
&&\!\!\!\!\!\!\!\!\!
T_G=(\mathcal{K}^{a_{1}}_{\,\,\,\,ea}\mathcal{K}^{ea_{2}}_{\,\,\,\,\,\,\,b}
\mathcal{K}^{a_{3}}_{\,\,\,\,fc}\mathcal{K}^{fa_{4}}_{\,\,\,\,\,\,\,d}
-2\mathcal{K}^{a_{1}\!a_{2}}_{\,\,\,\,\,\,\,\,\,\,a}\mathcal{K}^{a_{3}}_{
\,\,\,\,\,eb}\mathcal{K}^{e}_{\,\,fc}\mathcal{K}^{fa_{4}}_{\,\,\,\,\,\,\,\,d}
\nn\\
&& \ \ \ \
\,+2\mathcal{K}^{a_{1}\!a_{2}}_{\,\,\,\,\,\,\,\,\,\,a}\mathcal{K}^{a_{3}}_{
\,\,\,\,\,eb}\mathcal{K}^{ea_{4}}_{\,\,\,\,\,\,\,f}\mathcal{K}^{f}_{\,\,\,cd}
\nn\\
&& \ \ \ \
\,+2\mathcal{K}^{a_{1}\!a_{2}}_{\,\,\,\,\,\,\,\,\,\,a}\mathcal{K}^{a_{3}}_{
\,\,\,\,\,eb}\mathcal{K}^{ea_{4}}_{\,\,\,\,\,\,\,c,d})\delta^{\,a\,b\,c\,d}_{
a_{1}a_{2}a_{3}a_{4}}\,,
\label{TG}
\end{eqnarray}
with the generalized $\delta$ being the determinant of the Kronecker
deltas. Thus, $T_G$ is the teleparallel equivalent of $\bar{G}$, in the sense
that the action
\begin{equation}
S_{tel}^{(2)}
=\frac{1}{2\kappa_{D}^{2}}\int_{M}\!\!d^{D}\!x\,
e\,T_{G}\,,
\label{teleaction2}
\end{equation}
varied in terms of the vielbein gives exactly the same equations with the
action
\begin{equation}
S_{GB}
=\frac{1}{2\kappa_{D}^{2}}\int_{M}\!\!d^{D}\!x\,
e\,\bar{G}\,,
\label{GBaction2}
\end{equation}
varied in terms of the metric.

Having constructed the teleparallel equivalent of curvature invariants, one
can be based on them in order to build modified gravitational theories.
Thus, one can start from an action where $T$ is generalized to $F(T)$,
resulting to the so-called $F(T)$ gravity
\cite{Ferraro:2006jd,Linder:2010py,Chen:2010va,Myrzakulov:2010vz,Li:2011wu,Wu:2010mn,Wang:2011xf}.
Similarly, one can extend $T_G$ to $F(T_G)$ in the action, and since  $T_G$
is quartic in torsion then $F(T_G)$ cannot arise from any $F(T)$. Hence,
in \cite{Kofinas:2014owa} we combined both possible extensions and we
constructed the $F(T,T_G)$ modified gravity
\begin{equation}
S=\frac{1}{2\kappa_{D}^{2}}\!\int d^{D}\!x\,e\,F(T,T_G)\,,
\label{FTTGgravity}
\end{equation}
which is clearly different from
both $F(T)$ theory as well as from $F(R,G)$ gravity
\cite{Nojiri:2005jg,DeFelice:2008wz,Davis:2007ta}, and therefore it is
novel gravitational modification. We mention that TEGR (and therefore GR) is
obtained for $F(T,T_G)=-T$, while the usual Einstein-Gauss-Bonnet theory
arises for $F(T,T_G)=-T+\alpha T_G$, with $\alpha$
the Gauss-Bonnet coupling.

Let us now give the equations of motion in $D=4$ which is our main interest in the present paper.
Varying the action (\ref{FTTGgravity}) in terms of the vierbein, after various steps, we finally
obtain \cite{Kofinas:2014owa}
\begin{eqnarray}
&&\!\!\!\!\!\!
2(H^{[ac]b}\!+\!H^{[ba]c}\!-\!H^{[cb]a})_{,c}\!+\!2(H^{[ac]b}\!+\!H^{[ba]c}
\!-\!H^{[cb]a})C^{d}_{\,\,\,dc}\nn\\
&&\!\!\!\!\!\!\!\!+(2H^{[ac]d}\!+\!H^{dca})C^{b}_{\,\,\,cd}
\!+\!4H^{[db]c}C_{(dc)}^{\,\,\,\,\,\,\,\,a}\!+\!T^{a}_{\,\,\,cd}H^{cdb}-h^{ab
}\nn\\
&&\,\,\,\,\,\,\,\,\,\,\,\,\,\,\,\,\,\,\,\,\,\,\,\,\,\,\,\,\,\,\,\,\,\,\,\,\,
\,\,\,\,\,\,\,\,\,\,\,\,\,
+(F\!-\!TF_{T}\!-\!T_{G}F_{T_{G}})\eta^{ab}=0\,,
\label{genequations}
\end{eqnarray}
where
\begin{eqnarray}
&& \!\!\!\!\!\!\!
H^{abc}=F_{T}(\eta^{ac}\mathcal{K}^{bd}_{\,\,\,\,\,\,d}-\mathcal{K}^{bca})+F_
{T_{G}}\big[\nn\\
&&\!\!\!\!\!\!\!\epsilon^{cprt}\!\big(\!2\epsilon^{a}_{\,\,\,dkf}\mathcal{K}^
{bk}_{\,\,\,\,\,p}
\mathcal{K}^{d}_{\,\,\,qr}
\!\!+\!\epsilon_{qdkf}\mathcal{K}^{ak}_{\,\,\,\,\,p}\mathcal{K}^{bd}_{\,\,\,\,\,\,r}\!\!+\!
\epsilon^{ab}_{\,\,\,\,\,\,kf}\mathcal{K}^{k}_{\,\,\,dp}\mathcal{K}^{d}_{\,\,
\,qr}\!\big)\!
\mathcal{K}^{qf}_{\,\,\,\,\,\,t}\nn\\
&&\,\,\,\,\,\,\,\,\,\,\,\,\,\,\,\,\,\,\,\,\,\,\,\,\,\,\,\,\,\,\,\,\,\,\,\,\,
 \,\,\,\,\,\,\,
+\epsilon^{cprt}\epsilon^{ab}_{\,\,\,\,\,\,kd}\mathcal{K}^{fd}_{\,\,\,\,\,\,p
}
\big(\mathcal{K}^{k}_{\,\,fr,t}\!-\!\frac{1}{2}\mathcal{K}^{k}_{\,\,fq}C^{q}_
{\,\,\,tr}\big)\nn\\
&&\,\,\,\,\,\,\,\,\,\,\,\,\,\,\,\,\,\,\,\,\,\,\,\,\,\,\,\,\,\,\,\,\,\,\,\,\,
\,\,\,\,\,\,
+\epsilon^{cprt}\epsilon^{ak}_{\,\,\,\,\,\,df}\mathcal{K}^{df}_{\,\,\,\,p}
\big(\mathcal{K}^{b}_{\,\,kr,t}\!-\!\frac{1}{2}\mathcal{K}^{b}_{\,\,kq}C^{q}_
{\,\,\,tr}\big)\big]
\nn\\
&&\!\!+\epsilon^{cprt}\epsilon^{a}_{\,\,\,kdf}\Big[
\big(F_{T_{G}}\mathcal{K}^{bk}_{\,\,\,\,\,p}
\mathcal{K}^{df}_{\,\,\,\,\,r}\big)_{,t}\!+\!
F_{T_{G}}C^{q}_{\,\,\,pt}\mathcal{K}^{bk}_{\,\,\,\,\,[q}\mathcal{K}^{df}_{\,\,\,\,\,r]}\Big]\label{Habc22}
\end{eqnarray}
and
\begin{equation}
h^{ab}=F_{T}\epsilon^{a}_{\,\,\,kcd}\epsilon^{bpqd}\mathcal{K}^{k}_{\,\,\,fp}
\mathcal{K}^{fc}_{\,\,\,\,\,\,q}\,.
\label{hab22}
\end{equation}
We have used the notation $F_{T}=\partial F/\partial T$, $F_{T_{G}}=\partial F/\partial T_{G}$,
the (anti)symmetrization symbol contains the factor $1/2$, while the antisymmetric symbol $\epsilon_{abcd}$
has $\epsilon_{1234}=1$, $\epsilon^{1234}=-1$.

We close this section by making some comments on  $F(T,T_G)$ modified gravity itself, before proceeding to its cosmological
investigation. The first has to do with the Lorentz violation. In particular, as we discussed also in \cite{Kofinas:2014owa},
under the use
of the  Weitzenb{\"{o}}ck connection the torsion scalar $T$ remains diffeomorphism invariant,
however the Lorentz invariance has been lost since we have chosen a specific
class of frames, namely the autoparallel orthonormal frames. Nevertheless, the equations of motion of the Lagrangian
$eT$, being the Einstein equations,
are still Lorentz covariant. On the contrary, when we replace  $T$  by a general function $f(T)$ in the action,
the new equations of motion  will not be covariant under Lorentz rotations of the vielbein, although they will indeed be form-invariant, and the
same features appear in the $F(T,T_G)$ extension. However, this is not a deficit (it is a sort of analogue of gauge fixing in gauge theories),
and the theory, although not Lorentz covariant, is meaningful. Definitely, not all vielbeins will be solutions of
the equations of motion, but those which solve the
equations will determine the metric uniquely.

The second comment is related to possible acausalities  and problems with the
Cauchy development of a constant-time hypersurface. Indeed, there are works claiming that a departure from
TEGR, as for instance in $f(T)$ gravity, with the subsequent local Lorentz violation, will lead to the above problems
\cite{Ong:2013qja}. In order to examine whether one also has these problems in the present scenario of $F(T,T_G)$ gravity,
he would need to perform a very complicated analysis,
extending the characteristics method of \cite{Ong:2013qja} for this case, although at first sight one does expect to indeed find them.
Nevertheless, even if this proves to be the case, it does not
mean that the theory has to be ruled out, since one could still handle $f(T)$ gravity (and similarly $F(T,T_G)$ one) as an effective theory,
in the regime of validity of
which the extra degrees of freedom can be removed or be excited in a
healthy way (alternatively one could reformulate the theory using Lagrange multipliers) \cite{Ong:2013qja}.
However, there is a possibility that these
problems might be related to the restricting use of the  Weitzenb{\"{o}}ck connection, since the formulation of TEGR and its modifications using other
connections (still in the ``teleparallel class'') does not seem to be problematic, and thus, the general formulation of  $F(T,T_G)$ gravity that was presented in
 \cite{Kofinas:2014owa} might be free of the above disadvantages. These issues definitely need
further investigation, and the discussion is still open in the literature.

\section{$F(T,T_G)$ cosmology}
\label{Fcosmo}

In this section we apply $F(T,T_G)$ gravity in a cosmological
framework. Firstly, we add the matter sector along the gravitational one,
that is we start by the total action
\begin{eqnarray}
S_{tot} =\frac{1}{2\kappa^{2}}\!\int d^{4}\!x\,e\,F(T,T_G)\,+S_m\,,
\label{fGBtelaction}
\end{eqnarray}
where $S_m$ corresponds to a matter energy-momentum tensor
$\Theta^{\mu\nu}$ and  $\kappa^2= 8\pi G$ is the four-dimensional
Newton's constant.
Secondly, in order to investigate the cosmological implications of the above
action, we
consider  a spatially flat cosmological ansatz
\begin{equation}
ds^{2}=-N^2(t)
dt^{2}+a^{2}(t)\delta_{\hat{i}\hat{j}}dx^{\hat{i}}dx^{\hat{j}}\,,
\label{metriccosmo}
\end{equation}
where $a(t)$ is the scale factor and $N(t)$ is the lapse function (the
hat indices run in the three spatial coordinates).
This metric arises from the diagonal vierbein
\begin{equation}
\label{vierbeincosmo}
e^{a}_{\,\,\,\mu}=\text{diag}(N(t),a(t),a(t),a(t))
\end{equation}
through (\ref{metrdef}),
while the dual vierbein is
$e_{a}^{\,\,\,\mu}=\text{diag}(N^{-1}(t),a^{-1}(t),
a^{-1}(t),a^{-1}(t))$, and its determinant $e=N(t)a(t)^{3}$.

Considering as
usual $N(t)=1$ and inserting the vierbein (\ref{vierbeincosmo}) into
relations (\ref{Tscalar}) and  (\ref{TG}), we find
\begin{eqnarray}
\label{Tcosmo}
&& \!\!\!\!\! \!\!\!\!\!\!\!\!\!\!\!\!\!T=6\frac{\dot{a}^{2}}{
a^{2}}=6H^2\\
&&\!\!\!\!\!\!\!\!\!\!\!\!\!\!\!\!\!\!T_G=24\frac{\dot{a}^2}{a^2}
\,\frac{\ddot{a}}{a}=24H^2\big(\dot{H}+H^2\big),
 \label{TGcosmo}
\end{eqnarray}
where $H=\frac{\dot{a}}{a}$ is the Hubble parameter and dots denote
differentiation with respect to $t$. Additionally, inserting
(\ref{vierbeincosmo}) into the general equations of motion
(\ref{genequations}), after some algebra we obtain the Friedmann
equations
\begin{equation}
F-12H^{2}F_{T}-T_{G}F_{T_{G}}+24H^{3}\dot{F_{T_{G}}}=2\kappa^{2}\rho
\label{eqmN}
\end{equation}
\begin{eqnarray}
&&\!\!\!\!\!\!\!\!\!\!F-4(\dot{H}+3H^{2})F_{T}-4H\dot{F_{T}}\nn\\
&& \!\!\!\!- T_{G}F_{T_{G}}+\frac{2}{3H}T_{G}
\dot{F_{T_{G}}}+8H^{2}\ddot{F_{T_{G}}}=-2\kappa^{2}p\,,
\label{eqma}
\end{eqnarray}
where the right hand sides arise from the independent variation of the
matter action, considering it to correspond to a perfect fluid with energy
density $\rho$ and pressure $p$ (it is $\Theta^{tt}=\rho$,
$\Theta^{\hat{i}\hat{j}}=\frac{p}
{a^{2}}\delta^{\hat{i}\hat{j}}$, $\Theta^{\hat{i}}_{\,\,\hat{i}}=3p$). In the above expressions
it is $\dot{F_{T}}=F_{TT}\dot{T}+F_{TT_{G}}\dot{T}_{G}$,
$\dot{F_{T_{G}}}=F_{TT_{G}}\dot{T}+F_{T_{G}T_{G}}\dot{T}_{G}$,
$\ddot{F_{T_{G}}}=F_{TTT_{G}}\dot{T}^{2}+2F_{TT_{G}T_{G}}\dot{T}
\dot{T}_{G}+F_{T_{G}T_{G}T_{G}}\dot{T}_{G}^{\,\,2}+
F_{TT_{G}}\ddot{T}+F_{T_{G}T_{G}}\ddot{T}_{G}$,
with $F_{TT}$, $F_{TT_{G}}$,\,... denoting multiple partial differentiations
of $F$ with respect to $T$, $T_{G}$.
Finally, $\dot{T}$, $\ddot{T}$ and $\dot{T}_{G}$, $\ddot{T}_{G}$ are obtained
by differentiating (\ref{Tcosmo}) and (\ref{TGcosmo}) respectively with
respect to time.

Let us make a comment here on the derivation of the above Friedmann
equations. We mention that we followed the robust way, that is we first
performed the general variation of the action resulting to the general
equations of motion (\ref{genequations}), and then we inserted the cosmological ansatz
(\ref{vierbeincosmo}), obtaining (\ref{eqmN}) and (\ref{eqma}). This
procedure in principle does not give the same results with the shortcut
procedure where one first inserts the cosmological
ansatz (\ref{vierbeincosmo}) in the action and then performs variation with
respect to $N$ and $a$, since variation and
ansatz-insertion do not commute in general, especially in theories with
higher-order derivatives \cite{Deser:2004yh,Weinberg:2008}. This shortcut method is a sort of minisuperspace
procedure since the (potential) additional degrees of freedom other than
those contained in the scale factor are frozen. However, as we
show in the Appendix, following the shortcut procedure in the cosmological
application of the scenario at hand leads exactly to the two Friedmann
equations (\ref{eqmN}) and (\ref{eqma}) of the robust procedure.

Setting $F(T,T_{G})=-T$ in equations (\ref{eqmN}), (\ref{eqma}), we
get the standard cosmological equations of General Relativity without a cosmological
constant. For $F(T,T_{G})=F(T)$ we get
\begin{equation}
F-12H^{2}F_{T}=2\kappa^{2}\rho \label{eqmN1}
\end{equation}
\begin{equation}
F-4(\dot{H}+3H^{2})F_{T}-4H\dot{F_{T}}=-2\kappa^{2}p\,,
\label{eqma1}
\end{equation}
which are recognized as the standard equations of $F(T)$ gravity. However,
note that due  to the various conventions adopted in the literature in the
definitions of the Riemann tensor, the torsion tensor and the Minkowski
metric, which may differ by a total sign, our function $F(T)$ may
correspond to $F(-T)$ in some other works (for instance
\cite{Linder:2010py}).

In order to parametrize the deviation of the theory $F(T,T_{G})$ from GR,  we
write $F(T,T_{G})=-T+f(T,T_{G})$.
Thus,  the modification of GR (for instance the effective dark energy
component) is included in the
function $f$. Equations (\ref{eqmN}), (\ref{eqma}) are then written as
\begin{equation}
6H^{2}+f-12H^{2}f_{T}-T_{G}f_{T_{G}}+24H^{3}\dot{f_{T_{G}}}=2\kappa^{2}\rho
\label{vgy}
\end{equation}
\begin{eqnarray}
&&\!\!\!\!
2(2\dot{H}+3H^{2})+f-4(\dot{H}+3H^{2})f_{T}-4H\dot{f_{T}}-T_{G}f_{T_{G}}\nn\\
&& \ \ \ \ \ \ \ \ \ \ \ \ \ \ \ \
+\frac{2}{3H}T_{G}\dot{f_{T_{G}}}+8H^{2}\ddot{f_{T_{G}}}=-2\kappa^{2}p\,.
\label{gyw}
\end{eqnarray}
The Friedmann equations (\ref{vgy}), (\ref{gyw}) can be rewritten in the
usual form
\begin{eqnarray}
\label{Fr1a}
H^{2}&=&\frac{\kappa^2}{3}(\rho+\rho_{DE})\\
\label{Fr2s}
\dot{H}&=&-\frac{\kappa^2}{2}(\rho+p+\rho_{DE}+p_{DE}),
\end{eqnarray}
where the energy density and pressure of the effective dark energy sector are
defined as
\begin{eqnarray}
&&\!\!\!\!\!\!\!\!\rho_{DE}=-\frac{1}{2\kappa^{2}}(f-12H^{2}f_{T}-T_{G}f_{T_{
G}}+24H^{3}\dot{f_{T_{G}}})
\label{hys}\\
&&\!\!\!\!\!\!\!\!
p_{DE}=\frac{1}{2\kappa^{2}}\Big[f-4(\dot{H}+3H^{2})f_{T}-4H\dot{f_{T}}-T_{G}
f_{T_{G}}\nn\\
&& \ \ \ \ \ \ \ \ \ \ \ \ \ \ \
+\frac{2}{3H}T_{G}\dot{f_{T_{G}}}+8H^{2}\ddot{f_{T_{G}}}\Big]\,.
\label{pwl}
\end{eqnarray}
In terms of the initial function $F$, we can write $\rho_{DE},p_{DE}$ as
\begin{eqnarray}
&&\!\!\!\!\!\!\!\!\!\!\!\!\!\!\!\!\!
\rho_{DE}\!=\!\frac{1}{2\kappa^{2}}(6H^{2}\!-\!F\!+\!12H^{2}F_{T}\!+\!T_{G}F_
{T_{G}}
\!-\!24H^{3}\dot{F_{T_{G}}})
\label{rhode}\\
&&\!\!\!\!\!\!\!\!\!\!\!\!\!\!\!\!\!
p_{DE}\!=\!\frac{1}{2\kappa^{2}}\Big[\!\!-\!2(2\dot{H}\!+\!3H^{2}
)\!+\!F\!-\!4(\dot{H}\!+\!3H^{2})F_{T}\nn\\
&& \ \ \ \ \ \ \,
-4H\dot{F_{T}}\!-\!T_{G}F_{T_{G}}\!+\!\frac{2}{3H}T_{G}\dot{F_{T_{G}}}
\!+\!8H^{2}\ddot{F_{T_{G}}}\!\!\Big]\!.
\label{pde}
\end{eqnarray}

Since the standard matter is conserved independently, $\dot{\rho}+3H(\rho+p)=0$,
we obtain from (\ref{hys}), (\ref{pwl}) that the dark energy density and
pressure also satisfy the usual evolution equation
\begin{eqnarray}
\dot{\rho}_{DE}+3H(\rho_{DE}+p_{DE})=0\,.
\end{eqnarray}
Finally, we can define the dark energy equation-of-state parameter as
\begin{eqnarray}
w_{DE}= \frac{p_{DE}}{\rho_{DE}}\,.
\end{eqnarray}

\section{Specific cases}
\label{speccases}

In the previous section we extracted the Friedmann equations of
general $F(T,T_{G})$ cosmology, and we defined the effective dark energy
sector. Thus, in this section we proceed to the investigation of some
specific $F(T,T_{G})$ cases, focusing on the evolution of observables such
as the various density parameters $\Omega_i=8\pi G\rho_i/(3H^2)$ and the
dark energy equation-of-state parameter $w_{DE}$.

\subsection{
$F(T,T_{G})=-T+\beta_{1}\sqrt{T^{2}+\beta_{2}T_{G}}+\alpha_{1}
T^{2}+\alpha_{2}T\sqrt{|T_{G}|}$}

Since $T_G$ contains quartic torsion terms, it will in general and
approximately be of
the same order with $T^2$. Therefore, $T$ and $\sqrt{T^{2}+\beta_{2}T_{G}}$
are of the same order, and thus, if one of them contributes during the
evolution the other will contribute too. Therefore, it would be very
interesting to consider modifications of the form
$F(T,T_{G})=-T+\beta_{1}\sqrt{T^{2}+\beta_{2}T_{G}}$, which are expected to
play an important role at late times. Note that the couplings $\beta_{1},\beta_{2}$ are dimensionless,
and so, no new mass scale enters at late times.
Nevertheless, in order to describe the early-times cosmology, one should
additionally include higher order corrections like
$T^{2}$. Since the scalar $T_{G}$ is of the same order with $T^{2}$, it
should be also included. However, since $T_{G}$ is topological
in four dimensions it cannot be included as it is, and therefore we use the
term $T\sqrt{|T_{G}|}$ which is also of the same order
with $T^{2}$ and non-trivial. Thus, the total function $F$ is taken to
be
\begin{equation}
F(T,T_{G})=-T+\beta_{1}\sqrt{T^{2}+\beta_{2}T_{G}}+
\alpha_{1} T^{2}+\alpha_{2}T\sqrt{|T_{G}|}\,.
\label{nhs}
\end{equation}
In summary, when the above function is used as an action, it gives rise
to a gravitational theory that can describe both inflation and late-times
acceleration in a unified way.

\begin{figure}[ht]
\includegraphics[scale=0.48]{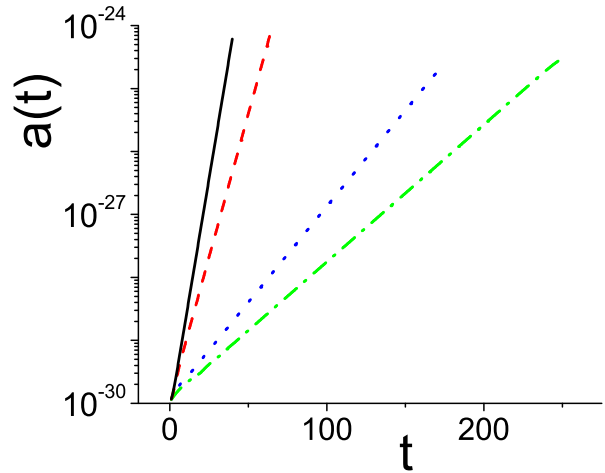}
\caption{
{\it{Four inflationary solutions for the   ansatz $F(T,T_{G})=\alpha_{1}
T^{2}+\alpha_{2}T\sqrt{|T_{G}|}-T+\beta_{1}\sqrt{T^{2}+\beta_{2}T_{G}}$,
corresponding to
a)  $\alpha_1=-2.8$, $\alpha_2=8$, $\beta_1=0.001$,  $\beta_2=1$
(black-solid),
b) $\alpha_1=-2$, $\alpha_2=8$, $\beta_1=0.001$, $\beta_2=1$ (red-dashed),
c) $\alpha_1=8$, $\alpha_2=8$, $\beta_1=0.001$, $\beta_2=1$ (blue-dotted),
d) $\alpha_1=20$, $\alpha_2=5$, $\beta_1=0.001$, $\beta_2=1$
(green-dashed-dotted).
All parameters are in Planck units.}} }
\label{Inflation}
\end{figure}

In order to examine the cosmological evolution of a universe governed by the
above unified action, we perform a numerical elaboration of the Friedman
equations (\ref{Fr1a}), (\ref{Fr2s}), with $\rho_{DE}$, $p_{DE}$ given by
equations
(\ref{rhode}), (\ref{pde}), under
the ansatz (\ref{nhs}). In Fig. \ref{Inflation} we present the
early-times, inflationary solutions for four parameter choices.
As we observe, inflationary, de-Sitter exponential expansions can be easily
obtained (with the exponent of the expansion determined by the model
parameters), although there is not an explicit cosmological constant term in
the
action, which is an advantage of the scenario. This was expected, since one
can easily extract analytical solutions of the Friedmann equations
(\ref{Fr1a}), (\ref{Fr2s}) with
$H\approx$ const (in which case $T$ and $T_G$ as also constants).

\begin{figure}[!]
\includegraphics[scale=0.48]{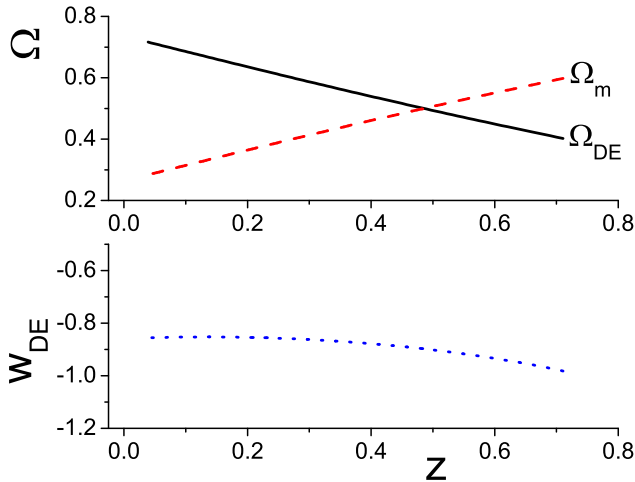}
\caption{ {\it{Upper graph: The evolution of the dark energy density
parameter $\Omega_{DE}$ (black-solid)  and the matter density
parameter $\Omega_{m}$ (red-dashed), as a function of the redshift $z$,
for the ansatz $F(T,T_{G})=\alpha_{1}
T^{2}+\alpha_{2}T\sqrt{|T_{G}|}-T+\beta_{1}\sqrt{T^{2}+\beta_{2}T_{G}}$
with $\alpha_1=0.001$, $\alpha_2=0.001$, $\beta_1=2.5$,
$\beta_2=1.5$.  Lower graph: The evolution of the corresponding dark
energy equation-of-state parameter $w_{DE}$. All parameters are in
units where the present Hubble parameter is $H_0=1$, and we have
imposed $\Omega_{m0}\approx0.3$, $\Omega_{DE0}\approx0.7$  at
present.}}} \label{Darkenergy}
\end{figure}

Let us now focus on the late-times evolution. In Fig. \ref{Darkenergy}
we depict the evolution of the matter and effective dark energy density
parameters, as well as the behavior of the dark energy equation-of-state
parameter, for a specific choice of the model parameters.
\begin{figure}[!]
\includegraphics[scale=0.48]{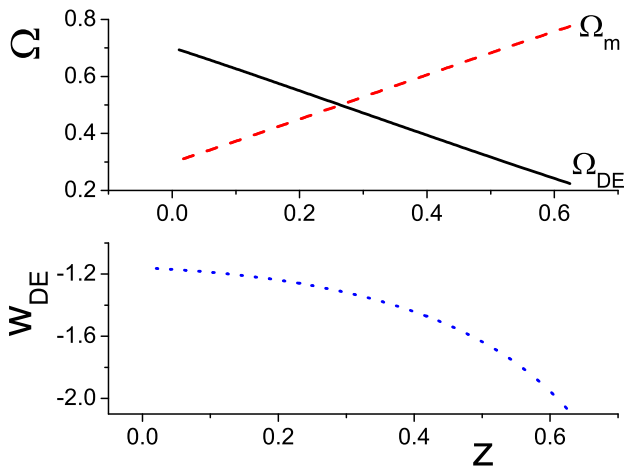}
\caption{
{\it{Upper graph: The evolution of the dark energy density parameter
$\Omega_{DE}$ (black-solid) and the matter density parameter
$\Omega_{m}$ (red-dashed), as a
function of the redshift $z$, for the   ansatz $F(T,T_{G})=
\alpha_{1}
T^{2}+\alpha_{2}T\sqrt{|T_{G}|}-T+\beta_{1}\sqrt{T^{2}+\beta_{2}T_{G}}$,
with $\alpha_1=0.001$, $\alpha_2=0.001$, $\beta_1=2.6$, $\beta_2=2$. Lower
graph: The
evolution of the corresponding dark energy equation-of-state parameter
$w_{DE}$. All parameters are in units where the present
Hubble parameter is $H_0=1$,
and we have imposed $\Omega_{m0}\approx0.3$, $\Omega_{DE0}\approx0.7$  at
present. }}}
\label{Darkenergy.phantom}
\end{figure}

As we see, we can obtain the observed behavior, where $\Omega_m$ decreases,
resulting to its current value of $\Omega_{m0}\approx0.3$, while
$\Omega_{DE}=1-\Omega_m$ increases. Concerning $w_{DE}$, we can see that in
this example it lies in the quintessence regime.

However, as it is usual in modified gravity \cite{Nojiri:2013ru}, the model
at hand can describe the phantom regime too, for a
region of the parameter space, which is an additional advantage.
In particular, in Fig. \ref{Darkenergy.phantom} we depict the
cosmological evolution for a parameter choice that leads $w_{DE}$ to the
phantom regime, while the density parameters maintain their observed
behavior. Similarly, note that the scenario can also exhibit the
phantom-divide crossing \cite{Cai:2009zp} too. Finally, note that one can
use dynamical-systems methods in order to examine in a systematic way the
late-times cosmological behavior of the scenario at hand, independently of the
initial conditions of the universe \cite{Kofinas:2014aka}.

\subsection{  $F(T,T_G)=-T+f(T^2+\beta_{2}T_G)$}

One can go beyond the simple model of the previous paragraph. In particular,
since as we already mentioned $T_G$ contains quartic torsion terms, it will
in general and approximately  be of the same order with $T^2$. Therefore,  it
would be interesting to consider modifications of the form
$F(T,T_G)=-T+f(T^2+\beta_{2}T_G)$. The involved building block is an
extension of the simple $T$, and thus, it can significantly  improve the
detailed cosmological behavior of a suitable reconstructed $F(T)$.
The general equations (\ref{vgy}), (\ref{gyw}) in this case become
\begin{eqnarray}
&&\!\!\!\!\!\!\!
6H^{2}+f-(24H^{2}T\!+\!\beta_{2}T_{G})f'+24\beta_{2}H^{3}(2T\dot{T}\!+\!\beta_{2}\dot{T}_{G})f''\nn\\
&&\quad\quad\quad\quad\quad\quad\quad\quad\quad\quad\quad\quad\quad\quad\quad\quad\quad=2\kappa^{2}\rho
\label{Fr1a22}
\end{eqnarray}
\begin{eqnarray}
&&\!\!\!\!\!\!
2(2\dot{H}+3H^{2})+f-[8(\dot{H}\!+\!3H^{2})T\!+\!8H\dot{T}\!+\!\beta_{2}T_{G}]f'\nn\\
&&\!\!\!\!\!\!+\Big{\{}\!\Big[\frac{2\beta_{2}T_{G}}{3H}\!-\!8HT\!\Big](2T\dot{T}\!\!+\!\!\beta_{2}\dot{T}_{G})
\!+\!8\beta_{2}H^{2}(\!2T\dot{T}\!\!+\!\!\beta_{2}\dot{T}_{G})^{^{\!\cdot}}\!\Big{\}}f''\nn\\
&&\quad\quad\quad\quad\quad
+8\beta_{2}H^{2}(2T\dot{T}+\beta_{2}\dot{T}_{G})^{2}f'''=-2\kappa^{2}p\,, \label{Fr2s22}
\end{eqnarray}
where $f',f'',f'''$ denote the derivatives of the function $f$ and are
evaluated at $T^{2}+\beta_{2}T_{G}$.

\begin{figure}[ht]
\includegraphics[scale=0.48]{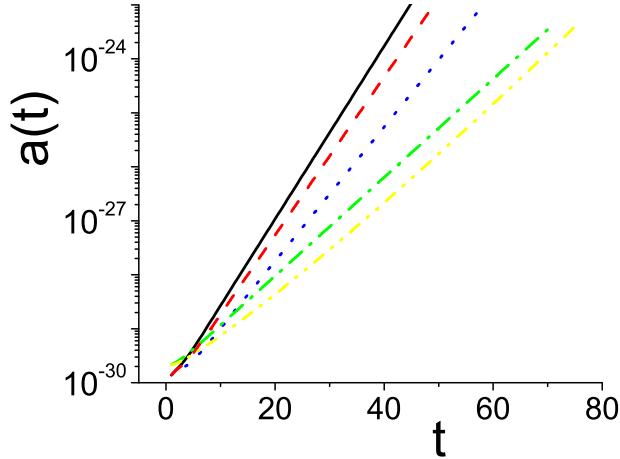}
\caption{
{\it{Five inflationary solutions for the   ansatz
$F(T,T_{G})=-T+\beta_1(T^2+\beta_{2}T_G)+\beta_3(T^2+\beta_{4}T_G)^2$,
corresponding to
a)  $\beta_1=-0.01$, $\beta_2=1$, $\beta_3=-1$,  $\beta_4=-2$ (black-solid),
b)  $\beta_1=-0.1$, $\beta_2=1$, $\beta_3=-2$,  $\beta_4=-2$ (red-dashed),
c)  $\beta_1=-0.01$, $\beta_2=1$, $\beta_3=-1$,  $\beta_4=-5$ (blue-dotted),
d)  $\beta_1=-0.01$, $\beta_2=1$, $\beta_3=-6$,  $\beta_4=-6$
(green-dashed-dotted),
e)  $\beta_1=-0.01$, $\beta_2=1$, $\beta_3=-10$,  $\beta_4=-10$
(yellow-dashed-dotted-dotted).
All parameters are in Planck units.}} }
\label{Inflationnewmodel}
\end{figure}

As a representative example we choose the case
$F(T,T_{G})=-T+\beta_1(T^2+\beta_{2}T_G)+\beta_3(T^2+\beta_{4}T_G)^2$, that
is keeping up to fourth-order torsion terms (one could easily proceed to the
investigation of other ansatzes in this
class and to the detailed description of a unified picture of inflation and
late-times acceleration). As expected, the higher-order
torsion terms are significant at early times, and thus they can easily drive
inflation. In Fig. \ref{Inflationnewmodel} we show the
early-times, inflationary solutions for five parameter choices, changing in
particular the value of $\beta_4$ in order to see the effect of $T_G$ on the
evolution (the value of $\beta_2$ is irrelevant since the linear $T_G$ term does
not have any effect, since $T_G$, similarly to $G$, is topological
invariant).

As we observe, inflationary evolution, that is de-Sitter exponential
expansions can be easily obtained, and the expansion-exponent is determined
by the model parameters. The significant advantage is that the exponential
expansion is obtained without an explicit cosmological constant term in the
action. Again, this was expected since we can easily extract analytical
solutions of the Friedmann equations (\ref{Fr1a22}), (\ref{Fr2s22}) with
$H\approx$ const (in which case $T$ and $T_G$ as also constants).
Additionally, note that in this case the inflation realization is more
efficient comparing to the model of the previous subsection, since it leads
to more e-foldings in less time, as expected since now higher-order torsion
terms are considered.

In the above analysis we showed that $f(T,T_G)$ cosmology can be very efficient in describing the
evolution of the universe at the background level. However, before considering any cosmological model as a candidate for
the description of nature it is necessary to perform a detailed investigation of its perturbations, namely to examine whether
the obtained solutions are stable. Furthermore, especially in theories with local Lorentz invariance violation, new degrees of freedom are introduced,
the behavior of which is not guaranteed that is stable (for instance this is the case in  the initial version of Ho\v{r}ava-Lifshitz gravity
\cite{Bogdanos:2009uj}, in the initial version of de Rham-Gabadadze-Tolley massive gravity \cite{DeFelice:2012mx}, etc), and
this makes the perturbation analysis of such theories even more imperative.
Although such a detailed and complete analysis of
the cosmological perturbations of $f(T,T_G)$ gravity is necessary, its various complications and lengthy calculations
make it more convenient to be examined in a separate project \cite{usinprep}. However, for the moment we would like to mention that in the case of simple
 $f(T)$ gravity, the perturbations of which have been examined in detail  \cite{Chen:2010va,Li:2011wu},
 one does obtain instabilities, but there are many classes of $f(T)$ ansantzes and/or parameter-space regions, where the perturbations are well-behaved. This is
 a good indication that we could  expect to find a similar behavior in
 $f(T,T_G)$ gravity too, although we need to indeed verify this under the detailed perturbation analysis.

\section{Conclusions}
\label{Conclusions}

In this work we investigated the cosmological applications of $F(T,T_G)$
gravity, which is a modified gravity based on the torsion scalar $T$ and the
teleparallel equivalent of the Gauss-Bonnet combination $T_G$. $F(T,T_G)$
gravity is different from both the simple $F(T)$ theory, as well as from the
curvature modification $F(R,G)$, and thus it is a novel class of
gravitational modification. First, we extracted the general Friedmann
equations and we defined the effective dark energy sector consisting of
torsional combinations. Then, choosing specific $F(T,T_G)$ ansatzes we
performed a detailed study of various observables, such as the matter and
dark energy density parameters and the dark energy equation-of-state
parameters.

Amongst the huge number of possible ansatzes, an interesting option is the
construction of terms of the same order as $T$ or $T^2$ using $T_G$, for
instance new combinations of the form $T+\alpha \sqrt{|T_{G}|}$ or
$T^{2}+\beta T_{G}$ can participate in the $F(T,T_G)$ function.
The resulting cosmology leads to interesting behaviors.
Firstly, the scenario can describe the inflationary regime, without an
inflaton field. Secondly, at late times it provides an effective dark energy
sector which can drive the acceleration of the universe, along with the
correct evolution of the density parameters, without the need of a
cosmological constant. Furthermore, the dark energy equation-of-state
parameter can be quintessence or phantom-like, or experiences the
phantom-divide crossing, depending on the parameters of the model. Another
possible ansatz is the  $F(T,T_G)=-T+f(T^2+\beta_{2}T_G)$, the simplest
application of which can also easily lead to inflationary behavior. These
features make the proposed modified gravity a good candidate for the
description of Nature.

\begin{acknowledgments}
The research of ENS is implemented within the framework of the
Operational Program ``Education and Lifelong Learning'' (Actions
Beneficiary: General Secretariat for Research and Technology), and
is co-financed by the European Social Fund (ESF) and the Greek State.
\end{acknowledgments}

\appendix

\section{Shortcut procedure for the extraction of the field equations of
$F(T,T_G)$ cosmology} \label{Fieldequations}

In this Appendix we follow the shortcut procedure in order to derive the
field equations of $F(T,T_G)$ cosmology. In particular, instead of
performing the variation of the action (\ref{fGBtelaction}) in terms of the general vierbein,
obtaining the general field
equations (\ref{genequations}), and then insert into them the cosmological
vierbein ansatz (\ref{vierbeincosmo}), we will first insert the cosmological
vierbein ansatz into the action and then perform the variation in terms of
the scale factor $a(t)$ and the lapse function $N(t)$. Although in principle
the shortcut procedure is not guaranteed that it will give the same results
as the first robust method \cite{Deser:2004yh,Weinberg:2008}, especially
when the action involves higher-order derivatives, in this specific example
it proves that we do obtain the same results indeed.

Under the cosmological ansatz (\ref{vierbeincosmo}), namely
\begin{equation}
\label{vierbeincosmo2}
e^{a}_{\,\,\,\mu}=\text{diag}(N(t),a(t),a(t),a(t)),
\end{equation}
the scalars $T$ and $T_G$ become
\begin{eqnarray}
\label{Tcosmo22}
&& \!\!\!\!\!\!\!\!\!\!\!\!\!\!\!\!\!T=6\frac{\dot{a}^{2}}{N^2
a^{2}}=6H^2\\
&& \!\!\!\!\!\!\!\!\!\!\!\!\!\!\!\!\! T_G=24\frac{\dot{a}^2}{
N^2a^2 }
\Big(\frac { \ddot {a }} { N^2 a}-\frac{\dot{N}\dot{a}}{N^3 a}
\Big)\nonumber\\
&& \!\!\!\!\!\!\!\! =24H^2\Big(\frac{\dot{H}}{N}
+H^2\Big).
 \label{TGcosmo22}
\end{eqnarray}
 Therefore, insertion into the total action (\ref{fGBtelaction}) gives
\begin{equation}
S_{tot}=\frac{1}{2\kappa^{2}}\int \!dt\,
Na^{3}\,F\Big(6H^2,\,24H^2\Big(\frac{\dot{H}}{N}+H^2\Big)\Big)\,+S_m \ .
\label{Scosmo2}
\end{equation}

Let us now perform the variation of (\ref{Scosmo2}) with respect to $N$
and $a$. Since $\delta_{N}H=-H\frac{\delta N}{N}$, we obtain
\begin{eqnarray}
&&\delta_{N}T=-12H^{2}\frac{\delta N}{N}\nonumber\\
&&\delta_{N}T_{G}
=-96H^{2}\Big[\Big(\frac{\dot{H}}{N}+H^{2}-\frac{H\dot{N}}{4N^{2}}
\Big)\frac{\delta N}{N}
+\frac{H}{4N^{2}}(\delta N)^{^{\centerdot}}\Big].\nonumber
\end{eqnarray}
Similarly, since
$\delta_{a}H=\frac{(\delta a)^{^{\centerdot}}}{Na}-
H\frac{\delta a}{a}$, we acquire
\begin{eqnarray}
&&\delta_{a}T=\frac{12H}{Na}\Big[(\delta
a)^{^{\centerdot}}-NH\delta a\Big]\nonumber\\
&&\delta_{a}T_{G}=\frac{24H}{Na}\Bigg[\frac{H}{N}(\delta
a)^{^{\centerdot\centerdot}}
+\Big(\frac{2\dot{H}}{N}+2H^{2}-\frac{H\dot{N}}{N^{2}}\Big)(\delta
a)^{^{\centerdot}}\nonumber\\
&&\ \ \ \ \ \ \ \ \ \ \ \ \ \ \ \ \ \ \ \ \ \ \
\ \ \ \ \ \ \
-3NH\Big(\frac{\dot{H}}{N}+H^{2}\Big)\delta a\Bigg].\nonumber
\end{eqnarray}
Therefore, variation of the gravitational part of the action (\ref{Scosmo2})
with respect to $N$ and $a$ gives
\begin{eqnarray}
&&\!\!\!\!\!\!\!\!\!\!\!
\delta_{N}S=\frac{1}{2\kappa^{2}}\int \!dt\,a^{3}\Bigg[F-12H^{2}
F_{T}+\frac{24}{a^ { 3 } } \Big(\frac{a^{3}H^{3}}{N}F_{T_{G}}
\Big)^{^{\!\!\centerdot}}\nonumber\\
&&\ \ \ \ \ \ \ \ \ \ \ \ \
-96H^{2}\Big(\frac{\dot{H}}{N}
+H^{2}-\frac{H\dot{N}}{4N^{2}}\Big)F_{T_{G}}\Bigg]\delta N
\label{varyN}
\end{eqnarray}
\begin{eqnarray}
&&\!\!\!\!\!\!\delta_{a}S\!=\!\frac{3}{2\kappa^{2}}\!\int\!dt\,
Na^{2}\Bigg\{\!F-4H^{2}F_{T}-24H^{2}\Big(\!\frac{\dot{H}}{N}
+H^{2}\!\Big)F_{T_{G}}\nonumber\\
&&\ \ \ \ \ \ \ \
-\frac{4}{a^{2}\!N}\Bigg[a^{2}\!HF_{T}+2a^{2}
\!H\Big(\!\frac{2\dot{H}}{N}+2H^{2}-\frac{H\dot{N}}{N^{2}}\!\Big)
F_{T_{G}}\!\Bigg]^{^{\!\centerdot}}\nonumber\\
&&\ \ \ \ \ \ \ \ \
+\frac{8}{a^{2}\!N}\Big(\!\frac{a^{2}\!H^{2}}{N}F_{T_{G}}\!\Big)^{^
{\!\!\centerdot\centerdot}}
\Bigg\}\delta a,
\label{varya}
\end{eqnarray}
where $F_{T}\equiv\partial F/\partial T$ and $F_{T_{G}}\equiv\partial
F/\partial T_{G}$.

Additionally, variation of  $S_{m}$ gives
\begin{equation}
\delta S_{m}=\frac{1}{2}\int \!d^{4}x \,e\,\Theta^{\mu\nu}\delta g_{\mu\nu},\nn
\end{equation}
and its time-dependent part is
\begin{equation}
\delta S_{m}=-\int \!dt \,N^{2}a^{3}\Theta^{tt}\delta N
+\int \!dt\, Na^{2}\Theta^{\hat{i}}_{\,\,\hat{i}}\delta a\,,
\end{equation}
where hat indices run in the three spatial coordinates.

In summary, taking into account the total action variation, and  setting as
usual $N=1$ in the end, the obtained field equations, that is the
Friedmann equations, take the form
\begin{eqnarray}
&&\!\!\!\!\!\!\!\!\!\!\!\!\!\!\!\!\!\!\!\!\!\!\!F-12H^{2}F_{T}-96H^{2}\big(
\dot{H}+H^{2} \big)F_{T_{G}}\nonumber\\
&&\ \ \ \ \ \ \
+\frac{24}{a^{3}}\big(a^{3}H^{3}F_{T_{G}}
\big)^{^{\!\centerdot}}=2\kappa^{2}\Theta^{tt}
\label{eqmN0}
\end{eqnarray}
\begin{eqnarray}
&&\!\!\!\!\!\!\!\!\!\!\!\!\!F-4H^{2}F_{T}-24H^{2}
\big(\dot{H}+H^{2}\big)F_{T_{G}}\nn\\
&&\!\!\!\!\!\!
-\frac{4}{a^{2}}\Big[a^{2} HF_{T}+4a^{2}
H\big(\dot{H}+H^{2}\big)F_{T_{G}} \Big]^{^{\centerdot}}\nn\\
&&\!\!\!\!\!\!
+\frac{8}{a^{2}}\big(a^{2}H^{2}F_{T_{G}} \big)^{^
{\!\!\centerdot\centerdot}}
=-\frac{2}{3}\kappa^{2}\Theta^{\hat{i}}_{\,\,\hat{i}}\,.
\label{eqma0}
\end{eqnarray}
Additionally, if we consider the matter energy-momentum tensor to correspond
to a perfect fluid of energy density $\rho$ and pressure $p$, we insert in the above
field equations $\Theta^{tt}=\rho$,
$\Theta^{\hat{i}\hat{j}}=\frac{p}
{a^{2}}\delta^{\hat{i}\hat{j}}$, $\Theta^{\hat{i}}_{\,\,\hat{i}}=3p$.

Lastly, we can re-organize the terms, performing the involved time
derivatives, resulting in the end to
\begin{equation}
F-12H^{2}F_{T}-T_{G}F_{T_{G}}+24H^{3}\dot{F_{T_{G}}}=2\kappa^{2}\rho
\label{eqmN22}
\end{equation}
\begin{eqnarray}
&&\!\!\!\!\!\!\!\!\!\!F-4(\dot{H}+3H^{2})F_{T}-4H\dot{F_{T}}\nn\\
&& \!\!\!\!- T_{G}F_{T_{G}}+\frac{2}{3H}T_{G}
\dot{F_{T_{G}}}+8H^{2}\ddot{F_{T_{G}}}=-2\kappa^{2}p\,,
\label{eqma22}
\end{eqnarray}
where $\dot{F_{T}}=F_{TT}\dot{T}+F_{TT_{G}}\dot{T}_{G}$,
$\dot{F_{T_{G}}}=F_{TT_{G}}\dot{T}+F_{T_{G}T_{G}}\dot{T}_{G}$,
$\ddot{F_{T_{G}}}=F_{TTT_{G}}\dot{T}^{2}+2F_{TT_{G}T_{G}}\dot{T}
\dot{T}_{G}+F_{T_{G}T_{G}T_{G}}\dot{T}_{G}^{\,\,2}+
F_{TT_{G}}\ddot{T}+F_{T_{G}T_{G}}\ddot{T}_{G}$,
with $F_{TT}$, $F_{TT_{G}}$,\,... denoting multiple partial differentiations
of $F$ with respect to $T$, $T_{G}$.
Here, $\dot{T}$, $\ddot{T}$ are obtained by differentiating $T=6H^{2}$ with
respect to time, while
$\dot{T}_{G}$, $\ddot{T}_{G}$ by differentiating
$T_{G}=24H^{2}(\dot{H}+H^{2})$.

As we observe, the two Friedmann equations (\ref{eqmN22}) and (\ref{eqma22}),
derived with the above shortcut procedure, coincide with the two Friedmann
equations (\ref{eqmN}) and (\ref{eqma}) derived with the robust procedure in
sections \ref{model} and \ref{Fcosmo}.


\end{document}